\documentclass[twocolumn,epsf,prd,floatfix,nofootinbib]{revtex4}
\usepackage{graphicx}
\usepackage{epsfig}
\usepackage{dcolumn}
\usepackage{bm}
\usepackage{latexsym}

\begin{document}

\title{UHE tau neutrino flux regeneration while skimming the Earth}
\author{Oscar Blanch Bigas$^1$, Olivier Deligny$^2$, K\'evin Payet$^3$, V\'eronique Van Elewyck$^{2,4}$}

\affiliation{
$^1$ LPNHE, CNRS/IN2P3 and  Universit\'e Paris VI-VII , Paris, France\\
$^2$ IPN, CNRS/IN2P3 and Universit\'e Paris Sud, Orsay, France\\
$^3$ LPSC, Universit\'e Joseph Fourier Grenoble 1, CNRS/IN2P3, INPG, Grenoble, France\\
$^4$ AstroParticules et Cosmologie (UMR 7165) and Universit\'e Paris 7, Paris, France}

\begin{abstract}
The detection of Earth-skimming tau neutrinos has turned into a very
promising strategy for the observation of ultrahigh-energy cosmic
neutrinos. The sensitivity of this channel crucially depends on the
parameters of the propagation of the tau neutrinos through the
terrestrial crust, which governs the flux of emerging tau leptons
that can be detected. One of the characteristics of this propagation
is the possibility of regeneration through multiple $\nu_\tau
\leftrightarrow \tau$ conversions, which are often neglected in the
standard picture. In this paper, we solve the transport equations
governing the $\nu_\tau$ propagation and compare the flux of
emerging tau leptons obtained allowing regeneration or not. We
discuss the validity of the approximation of neglecting the
 $\nu_\tau$ regeneration using different scenarios for the
 neutrino-nucleon cross sections and the tau energy losses.
\end{abstract}

\pacs{95.55.Vj, 13.60.-r, 14.60.Fg}

\maketitle

\section{Introduction}
\label{sec:intro}

With the advent of a new generation of large-scale detectors of
cosmic radiation, the observation of high-energy cosmic neutrinos
produced in distant astrophysical sites or possibly by other, more
exotic, mechanisms has become one of the major challenges of
astroparticle physics. In both astrophysical and exotic models,
substantial fluxes of electron and muon neutrinos are expected from
the disintegration of charged pions (and kaons) produced in the
interaction of accelerated particles with ambient matter and
radiation, either at the source location or between the source and the observer. Given the large distances traveled by these cosmic
neutrinos, approximately equal fluxes in $\nu_e , \nu_\mu$ and
$\nu_\tau$ are expected on Earth as a result of flavor mixing and
oscillations~\cite{nuoscil,Learned:1994wg}. Important efforts are
ongoing to build dedicated neutrino telescopes both in the
Southern~\cite{Andres:1999hm,ice3} and
Northern~\cite{antares,nemo,nestor,Aynutdinov:2005sc} hemisphere,
opening a window on the neutrino sky in the energy range $10^{-6}\
\mathrm{EeV} \leq E_\nu \leq 10^{-1}\ \mathrm{EeV}$. At even higher energies, other promising experiments are
developing the detection of coherent radio emission produced by
neutrino-induced showers in matter~\cite{anita,rice,forte,glue,hankins,james,swarup}.

Interestingly, recent studies have shown that the new generation of
ultrahigh-energy (UHE) cosmic ray detectors such as the Pierre
Auger Observatory~\cite{pao} and the HiRes Fly's Eye
detector~\cite{hires} have a comparable detection potential for
UHE neutrinos in the range of energies $10^{-1}\,\,
\mathrm{EeV} \leq E_\nu \leq 10^{2}\,\, \mathrm{EeV}$, where neutrinos are expected to be produced in the interaction of UHE
cosmic rays with the cosmic microwave background~\cite{cosmogenic}.
It is long known that downward-going showers induced by neutrinos
that penetrate deep in the atmosphere can in principle be identified
at large zenith angles ($\theta > 75^\circ$) where no background is
expected from hadronic primaries~\cite{beresmirnov,capelle}. It was
also pointed out more recently that the presence of $\nu_\tau$ in
the cosmic neutrino flux provides another promising channel of
detection for air shower detectors~\cite{antoine,fargion}.
Upward-going UHE tau neutrinos that graze the Earth just below the
horizon (often referred to as ``Earth-skimming neutrinos'') are
indeed likely to interact in the crust and produce a tau lepton
which may emerge and initiate an observable air shower, provided it
does not decay too far from the detector.

The sensitivity to such UHE Earth-skimming neutrinos crucially
depends on the conditions of the $\nu_\tau$ propagation through the
terrestrial crust, and on the correct estimation of the flux of $\tau$
leptons that emerge from the Earth. This propagation problem has
been widely discussed
in different contexts and with different
approximations~\cite{DRSS,bertou,feng,bottai,tseng,montaruli,aramo}. An
exhaustive treatment should account for $\tau$ and $\nu_\tau$
neutral-current (NC) and charged-current (CC) interactions with
nucleons, $\tau$ decay and energy losses. However the full coupled
transport equations admit no analytical solution, and even in the
case of Monte Carlo calculations, simplifications are usually made
such as dropping the $\tau$ weak interactions and neglecting
multiple regenerations of the $\nu_\tau$.

Such approximations are defendable in the standard case where the characteristic lengths for the $\tau$ CC interaction ($\sim 600$ km, at 1 EeV) and for the $\tau$ decay ($\sim 50$ km) are larger than that for energy losses ($\sim 6$ km). However, they might be challenged in other, more exotic scenarios. The knowledge of the neutrino cross section and the tau energy losses in the EeV energy range is indeed limited~\cite{sarkar, parente}, and these could be significantly affected at center-of-mass energies beyond the TeV by the onset of new physics
beyond the standard model. Several studies have even suggested that the comparison of the flux rates between down-going and Earth-skimming neutrinos could actually help constraining the neutrino properties at
ultrahigh energies, where no direct measurements
exist~\cite{irimia,anchordoqui}.

All these
considerations pinpoint the necessity of an accurate determination of the flux of emerging $\tau$ leptons. In this context, it is important to correctly describe all the contributions to the $\tau$
flux and to assess the impact of
simplifications in the description of the propagation problem.

The present paper therefore focuses on understanding the effects of
the $\nu_\tau$ and $\tau$ regeneration while skimming the Earth on the flux of emerging
$\tau$ leptons in different scenarios for the neutrino-nucleon
cross section and for the $\tau$ energy losses. In Sec.
\ref{sec:calcul},
 we present the transport equations for the $\nu$ and $\tau$ propagation,
 and the scenarios to be studied. In Sec. \ref{sec:results}, we compare the flux
 of emerging taus for different conditions of propagation. Finally, we
 present our conclusions in Sec. \ref{sec:conclusion}.

\section{$\nu_\tau$ propagation through the Earth: the general picture}
\label{sec:calcul}

\begin{figure}[!t]
    \centering
    \includegraphics[bbllx=1,bburx=548,bblly=350,bbury=490,width=9cm,clip=]{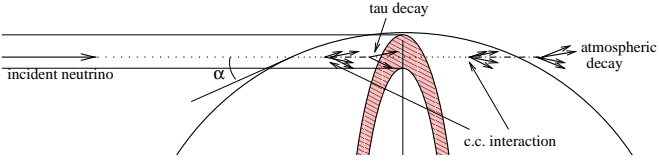}
   \caption{\small{Geometry of the transport problem.}}
    \label{fig:simul}
\end{figure}

The geometry of the propagation problem is described in
Fig.\ref{fig:simul}, where an example of regeneration chain through
multiple CC interactions and $\tau$ decays is sketched. Given a beam
of parallel neutrinos incident on the Earth at a given 
angle $\alpha$, the problem becomes unidimensional and the flux of
$\tau$ leptons that emerge only depends on the path length traveled
across the rock. In the hypothesis of a spherical Earth with a crust
of constant density, this length is directly related to $\alpha$,
 which is also the angle of the emerging tau.

\subsection{Transport equations}

We reproduce here a classical formulation in terms of the transport equations that
describe the
evolution of the $\nu_\tau$ and $\tau$ fluxes, $\Phi_{\nu_\tau}$ and $\Phi_\tau$,  along their path
through the Earth, accounting for all possible production and
absorption processes taking place within an infinitesimal
$\mathrm{d}x$:

\begin{eqnarray}
    &&\frac{\partial \Phi_{\nu_\tau}(E,x)}{\partial x} =
    - \frac{\Phi_{\nu_\tau}(E,x)}{\lambda_{\nu_\tau}^{CC}(E)}
    -\frac{\Phi_{\nu_\tau}(E,x)}{\lambda_{\nu_\tau}^{NC}(E)} \nonumber\\
    &&+ \rho\,\mathcal{N}_A\,\int\frac{\mathrm{d}y}{1-y}\Phi_{\nu_\tau}
    \bigg(\frac{E}{1-y},x\bigg)
    \frac{\mathrm{d}\sigma_{\nu_\tau}^{NC}(y,\frac{E}{1-y})}{\mathrm{d}y}\nonumber\\
    &&+ \rho\,\mathcal{N}_A\int\frac{\mathrm{d}y}{1-y}
    \Phi_{\tau}\bigg(\frac{E}{1-y},x\bigg)
    \frac{\mathrm{d}\sigma_{\tau}^{CC}(y,\frac{E}{1-y})}{\mathrm{d}y}\nonumber\\
    &&+ \frac{1}{c}\int\frac{\mathrm{d}y}{1-y}
    \Phi_{\tau}\bigg(\frac{E}{1-y},x\bigg)
    \frac{\mathrm{d}\Gamma_\tau (y,\frac{E}{1-y})}{\mathrm{d}y}
    \label{eq:PhiNuTau}
\end{eqnarray}

Here $\lambda_{\nu_\tau}^{NC}$ and $\lambda_{\nu_\tau}^{CC}$ are the
mean free paths corresponding respectively to NC and
CC interactions of the incident $\nu_\tau$, while
$\sigma_\tau^{CC}$ corresponds to the $\tau$ CC
interaction, which regenerates a $\nu_\tau$. $\Gamma_\tau$ is the
tau lepton lifetime.

\begin{figure}[!t]
    \centering
    \includegraphics[width=9cm]{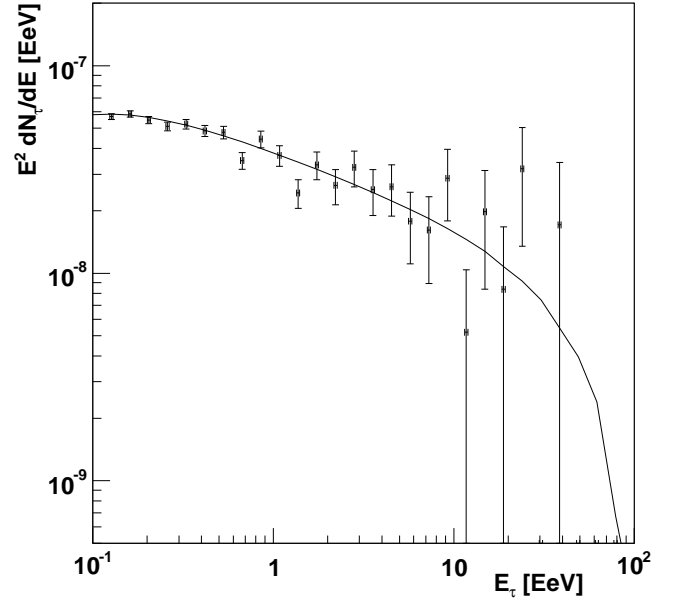}
    \caption{\small{Flux of emerging $\tau$'s (scaled by $E^2$) obtained by solving the transport Eqs.  (\ref{eq:PhiNuTau}) and (\ref{eq:PhiTau}), for incident tau neutrinos with an angle $\alpha$ up to 15$^{\circ}$ and injection flux $dN/dE = 4.6 \times 10^{7}$ E$_\nu^{-2}$~EeV$^{-1}$~sr$^{-1}$. The points correspond to the Monte Carlo solution, while the line is the solution obtained with the iterative method.}}
\label{fig:MCvsAna}
\end{figure}

The transport equation for the $\tau$ can be inferred
in a similar way. In this case, the small inelasticity of the $\tau$ radiative
interactions allows us to use the continuous energy loss approximation \cite{nous}.
Defining $\gamma(E)=-\rm{d}E/\rm{d}x$, the equation reads:
\begin{eqnarray}
    &&\frac{\partial \Phi_{\tau}(E,x)}{\partial x} =
    - \frac{\Phi_{\tau}(E,x)}{\lambda_\tau^{dec}(E)}
    -\frac{\Phi_{\tau}(E,x)}{\lambda_\tau^{CC}(E)}
    -\frac{\Phi_{\tau}(E,x)}{\lambda_\tau^{NC}(E)} \nonumber\\
    &&+\frac{\partial}{\partial E}(\gamma(E)\Phi_\tau(E,x)) \nonumber\\
    &&+\rho\,\mathcal{N}_A\,\int\frac{\mathrm{d}y}{1-y}
    \Phi_{\tau}\bigg(\frac{E}{1-y},x\bigg)
    \frac{\mathrm{d}\sigma_\tau^{NC}(y,\frac{E}{1-y})}{\mathrm{d}y}\nonumber\\
    &&+\rho\,\mathcal{N}_A\int\frac{\mathrm{d}y}{1-y}
    \Phi_{\nu_\tau}\bigg(\frac{E}{1-y},x\bigg)
    \frac{\mathrm{d}\sigma_{\nu_\tau}^{CC}(y,\frac{E}{1-y})}{\mathrm{d}y}
    \label{eq:PhiTau}
\end{eqnarray}
where $\lambda_{\tau}^{NC}$ and $\lambda_{\tau}^{CC}$ are now the
mean free paths associated with the  $\tau$ NC and CC interactions,
and $\lambda_\tau^{dec}$ is the decay length corresponding to
$\Gamma_{\tau}$.

We are thus left with a system of two coupled, integro-differential
equations, which cannot be solved analytically without making
simplifying assumptions on the nature and the respective importance
of the interactions that both the $\nu_\tau$ and the $\tau$ can
undergo. To preserve the generality of the solution we use a Monte
Carlo simulation that includes all the processes listed above and
follows the incident particle all the way through the rock, deciding
at each step on its fate according to the distributions encoded in
Eq.\ref{eq:PhiNuTau} and Eq.\ref{eq:PhiTau}.

An iterative method that also allows us to avoid simplifications was
proposed in~\cite{perrone}, and further discussed in~\cite{tseng} and
in~\cite{reya}. Our equations differ from those in~\cite{tseng,reya} in
that we also include the two terms describing $\tau$ NC interactions
in the transport equation for $\Phi_\tau$. Taking into account this
generalization, we have applied the iterative method (see the Appendix
 for details) to crosscheck the results from the Monte
Carlo simulation. In Fig.~\ref{fig:MCvsAna}, we show the good
agreement between both calculations for the flux of emerging $\tau$
leptons given an incident tau neutrino flux $dN/dE$ = 4.6$\cdot$10$^{7}$ E$_\nu^{-2}$~EeV$^{-1}$~sr$^{-1}$.

\subsection{Cross Sections}
\label{subsec:xs}

In the framework of the standard model, the neutrino-nucleon ($\nu N$) CC and
NC cross-sections, $\sigma_{\nu_\tau}^{CC}$ and $\sigma_{\nu_\tau}^{NC}$, describe deep-inelastic scattering processes. They
are expressed in terms of the structure functions of the nucleon,
which in turn depend on the individual parton distribution functions
(PDFs). The PDFs are obtained from measurements at accelerators, in
determinate ranges of Bjorken-$x$ and momentum transfer $Q^2$.
However, the range of parameters probed by the UHE neutrinos,
$x \sim 10^{-5}(1\,\, \mathrm{EeV}/E_\nu)$ and $Q^2$ up
to $\sim 10^{-5}\ \mathrm{ EeV}$, is outside the measured domain and therefore
extrapolations are needed.

We choose as a benchmark for the {\it standard} $\nu N$ cross section 
a recent parametrization presented in~\cite{sarkar}, which fits to updated HERA data:

\begin{eqnarray}
\label{eq:SarkarL} \sigma^{CC}_{std}(E) & = & 2.4\ \sigma^{NC}_{std}(E)   \nonumber\\
& = & 6.04\cdot \left(\frac{E}{10^{-9}\ \text{EeV}}\right)^{0.358} \text{pb}
\end{eqnarray}

As pointed out in the same paper, another, more speculative approach
based on the color glass condensate formalism~\cite{CGC} has also
been put forward recently to account for saturation at very low $x$. We will
also consider this case and use the following approximate parametrization
deduced from Fig. 1 of ~\cite{sarkar}:
\begin{eqnarray}
\label{eq:CGC} \hspace*{-0.5cm}\sigma^{CC}_{low}(E)&=&2.4\ \sigma^{NC}_{low}(E)  \nonumber\\
\hspace*{0.5cm}&=&3.89\cdot \left(\frac{E}{\text{EeV}}\right)^{0.170-0.037 \cdot \log_{10}\frac{E}{\text{EeV}}}  10^3 \text{pb}
\end{eqnarray}
as an example of {\it low} $\nu N$ cross section.

On the other hand, plenty of models using new physics predict an enhancement of the neutrino cross section~\cite{han}. As an example of {\it high} $\nu N$ cross section, we will use
\begin{equation}
\sigma^{CC}_{high}(E)= 2.4\ \sigma^{NC}_{high}(E) = 3\cdot \sigma^{CC}_{std}(E)
\end{equation}
without assuming any particular model.

\subsection{Tau energy losses}
\label{subsec:elosses}

The $\tau$ energy losses are usually parametrized using the
following formula:
\begin{equation}
\label{Elosses} \frac{dE_\tau}{dX} = - \alpha - \beta(E_\tau)\
E_\tau.
\end{equation}
where $\alpha$ is practically constant and accounts for ionization processes,
while $\beta(E_\tau)$ includes the radiative
contributions from bremsstrahlung (b), pair production (pp) and photonuclear
 (pn) interactions of the $\tau$.

For the purpose of our analysis, we adopt the same strategy as in
the previous section and extract from existing
computations~\cite{parente,Dutta:2005yt} a panel of ad-hoc
parametrizations of $\beta(E_\tau)$ in the relevant range of
energy, representative of the values found in the literature. The
set of parametrizations under study is the following (valid in the
range $10^{-1}$ EeV $\leq E_\nu \leq 10^{3}$ EeV):
\begin{eqnarray}
\beta_{std}(E_\tau) &=& \left(1.2+0.16\times
\ln(\frac{E_\tau}{10\,\text{EeV}})\right)\ 10^{-6}
\ \mathrm{g}^{-1}\mathrm{cm}^2 \nonumber \\
&& \nonumber\\
\beta_{high}(E_\tau) &=&1.36\
\left(\frac{E_\tau}{\text{EeV}}\right)^{0.35}\
10^{-6}\ \mathrm{g}^{-1}\mathrm{cm}^2\nonumber\\
&&\nonumber\\
\beta_{low}(E_\tau) &=&\left(0.3+0.06\times
\log_{10}(\frac{E_\tau}{\text{EeV}})\right)\ 10^{-6}\
\mathrm{g}^{-1}\mathrm{cm}^2\nonumber
\end{eqnarray}
respectively for a standard, a high and a low value of the energy
loss parameter $\beta$.

\section{The flux of emerging taus: results and comparisons}
\label{sec:results}

\begin{figure}[!t]
    \centering
    \includegraphics[width=9cm]{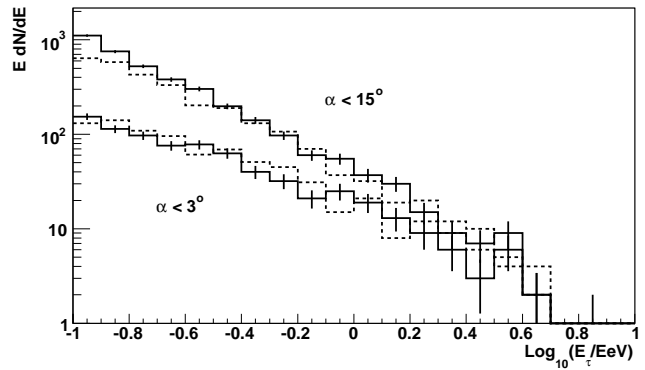}
    \caption{\small{Flux of emerging taus corresponding to incident tau neutrinos with
     an angle $\alpha$ up to 3$^{\circ}$ or 15$^{\circ}$ and $dN/dE = 4.6 \times 10^{7}~E_\nu^{-2}$~EeV$^{-1}$~sr$^{-1}$ using the standard combination of parametrizations $\sigma_{std}\otimes\beta_{std}$
    as defined in Secs.~\ref{subsec:xs} and~\ref{subsec:elosses}.
    The regeneration is neglected in the dashed histograms and included in the solid ones.}}
    \label{fig:StdE2_15Reg}
\end{figure}

\begin{figure}[!t]
    \centering
    \includegraphics[width=9cm]{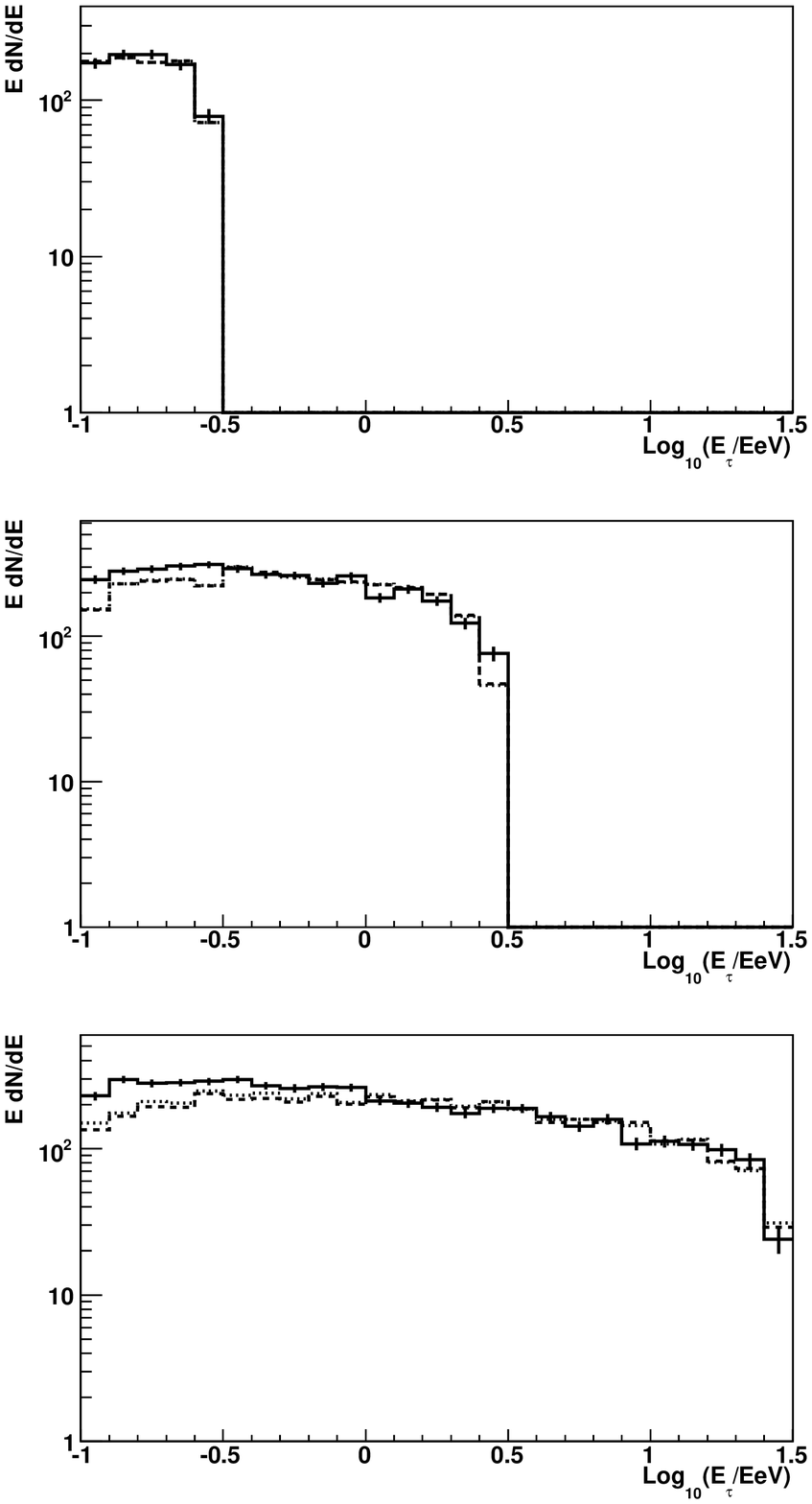}
    \caption{\small{Flux of emerging taus produced by incident  $\Phi_{\nu_\tau}(E_{\nu})$ = 4.6$\cdot$10$^8$ sr$^{-1}$ with energy $E_{\nu}$ = 0.3, 3 and 30 EeV (from top to bottom) and an angle up to 3$^\circ$. The standard combination of parametrizations $\sigma_{std}\otimes\beta_{std}$
    as defined in Secs.~\ref{subsec:xs} and~\ref{subsec:elosses} is used.
    The solid histogram takes into account all regeneration channels.
    The dotted histogram only accounts for the regeneration through the $\tau$ NC interaction.
    The dashed histogram neglects the regeneration.}}
    \label{fig:StdStdTau}
\end{figure}

\begin{figure}[!t]
    \centering
    \includegraphics[width=9cm]{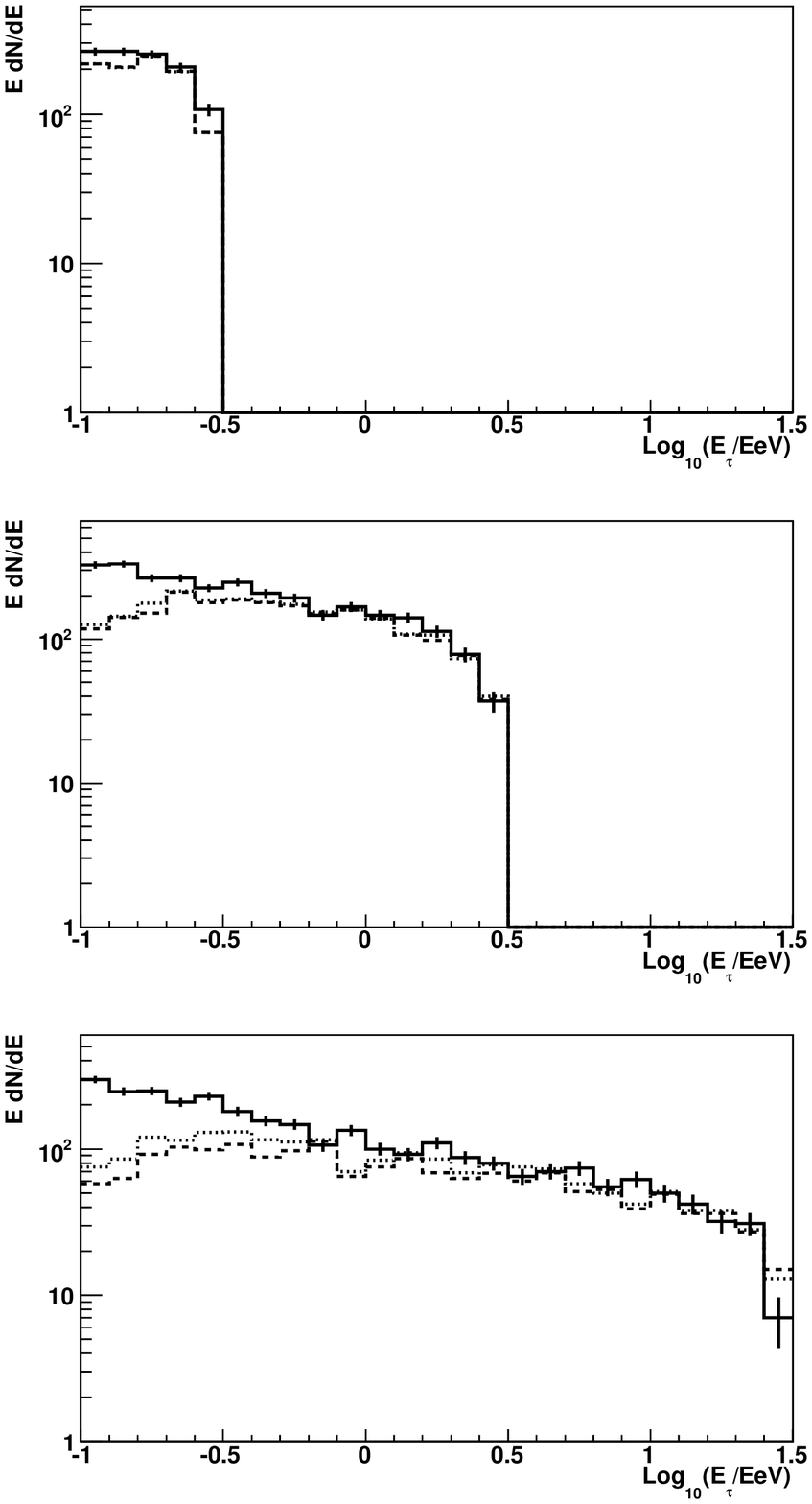}
    \caption{\small{Flux of emerging taus produced by incident  $\Phi_{\nu_\tau}(E_{\nu})$ = 4.6$\cdot$10$^8$ sr$^{-1}$ with energy $E_{\nu}$ = 0.3, 3 and 30 EeV (from top to bottom) and an angle up to 3$^\circ$. The combination $\sigma_{high}\otimes\beta_{std}$ as defined in Secs.~\ref{subsec:xs}
    and~\ref{subsec:elosses} is used (with the same histogram code as in Fig.~\ref{fig:StdStdTau}).}}
    \label{fig:StdHighTau}
\end{figure}

The Earth-skimming technique has been recently used to search for $\nu_{\tau}$ with the
HiRes telescopes~\cite{hires} and the Pierre Auger Observatory~\cite{pao}.  While both detectors have a similar energy threshold ($\sim$ 0.1 EeV), they  are sensitive to a different range in the emerging angle $\alpha$ of the $\tau$ leptons. The latter  
detects the particles from the air shower initiated by the $\tau$ decay that reach the ground and, hence, the detector will only be triggered by $\tau$ leptons slightly up-going ($\alpha < 3^{\circ}$). The former is designed to record the faint ultraviolet light emitted by nitrogen molecules that are excited as the shower traverses the atmosphere; it is therefore sensitive to $\tau$ leptons emerging with larger angles. 

In figure~\ref{fig:StdE2_15Reg}, we present the results obtained with the Monte Carlo approach for the
flux of emerging $\tau$ leptons with an angle $\alpha$ up to 3$^\circ$ and 15$^\circ$ produced by an incident tau neutrino flux $dN/dE$ = 4.6~10$^{7}~E_\nu^{-2}$~EeV$^{-1}$~sr$^{-1}$. All cross sections and energy loss parameters are here
set to {\it standard} values, as defined in Secs.~\ref{subsec:xs} and~\ref{subsec:elosses}. The effect of the regeneration is negligible if only almost horizontal emerging $\tau$ leptons ($\alpha < 3^\circ$) are taken into account. On the other hand, if the detector is sensitive to larger angles and energies down to a fraction of EeV, neglecting the regeneration becomes important already within this {\it standard} picture (30$\%$ less $\tau$ leptons at 0.3 EeV for an incident neutrino flux $dN/dE \propto E_\nu^{-2}$).

\begin{figure}[!t]
    \centering
    \includegraphics[width=9cm]{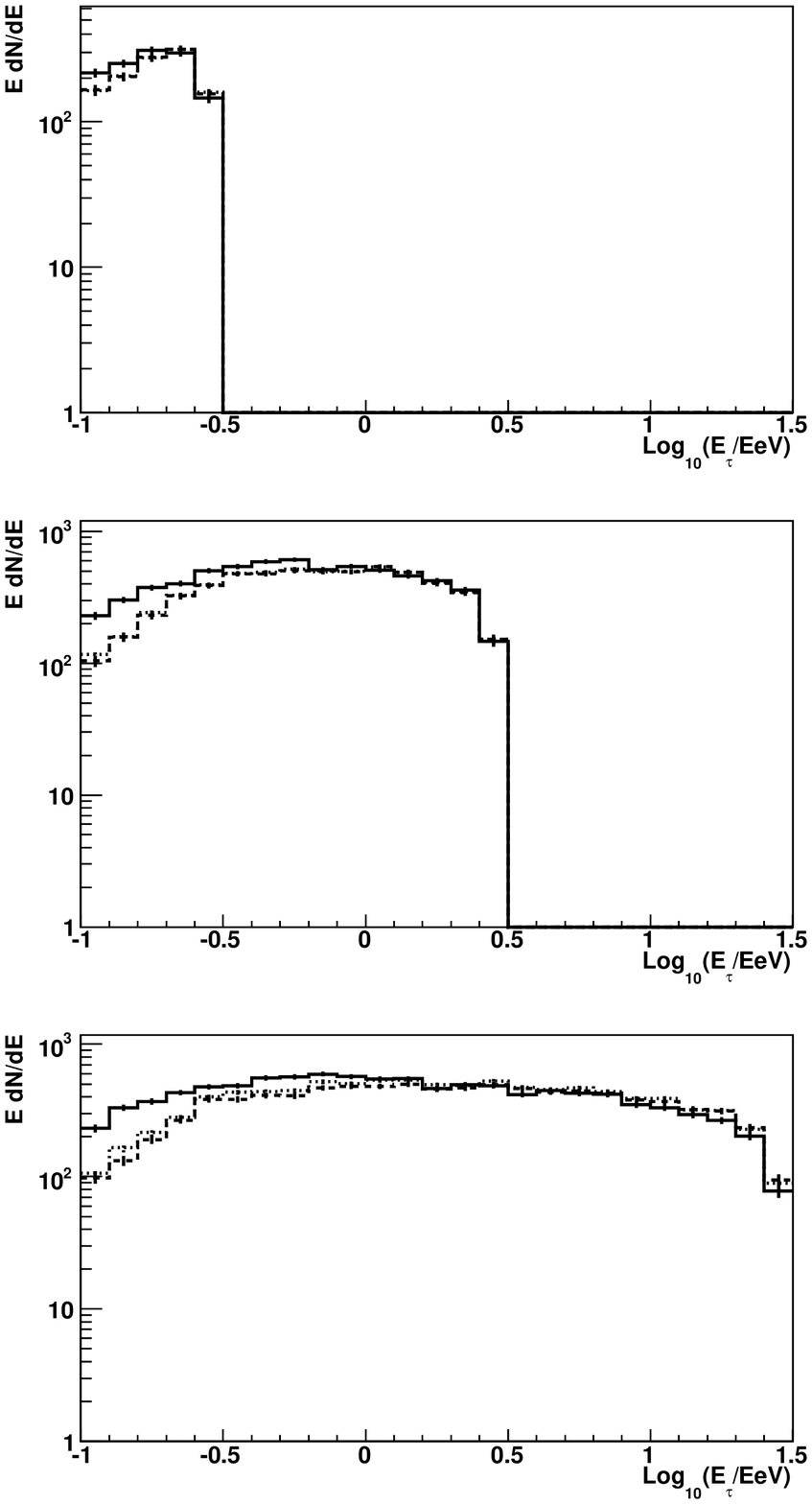}
    \caption{\small{Flux of emerging taus produced by incident  $\Phi_{\nu_\tau}(E_{\nu})$ = 4.6$\cdot$10$^8$ sr$^{-1}$ with energy $E_{\nu}$ = 0.3, 3 and 30 EeV (from top to bottom) and an angle up to 3$^\circ$. The combination $\sigma_{std}\otimes\beta_{low}$ as defined in Secs.~\ref{subsec:xs}and~\ref{subsec:elosses} is used (with the same histogram code as in Fig.~\ref{fig:StdStdTau}).}}
    \label{fig:LowStdTau}
\end{figure}

\begin{figure}[!t]
    \centering
    \includegraphics[width=9cm]{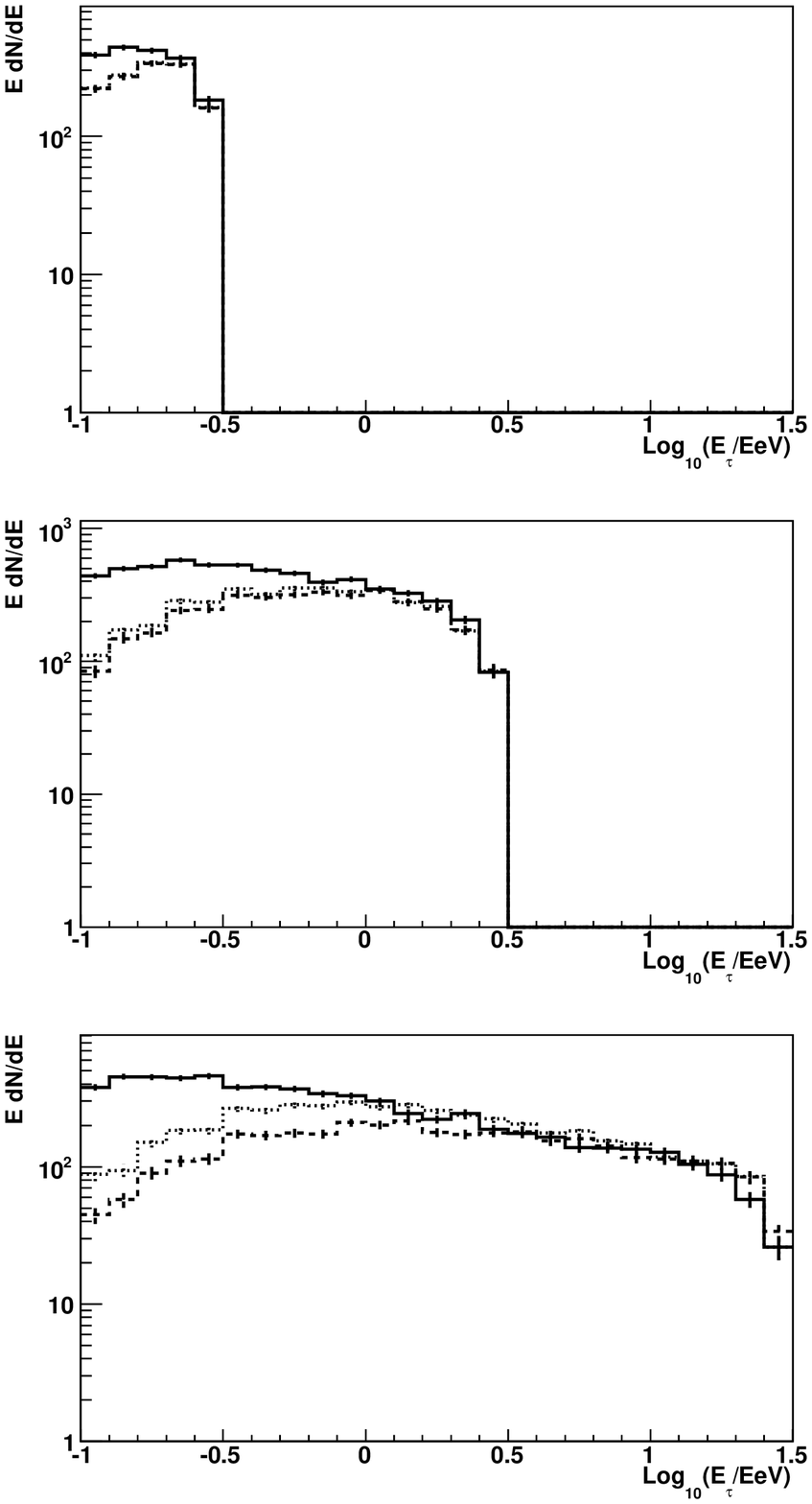}
    \caption{\small{Flux of emerging taus produced by incident  $\Phi_{\nu_\tau}(E_{\nu})$ = 4.6$\cdot$10$^8$ sr$^{-1}$ with energy $E_{\nu}$ = 0.3, 3 and 30 EeV (from top to bottom) and an angle up to 3$^\circ$. The combination $\sigma_{high}\otimes\beta_{low}$ as defined in Secs.~\ref{subsec:xs}
    and~\ref{subsec:elosses} is used (with the same histogram code as in Fig.~\ref{fig:StdStdTau}).}}
    \label{fig:LowHighTau}
\end{figure}

The approximation of neglecting the regeneration seems thus to be safe in the standard picture for detectors only sensitive to almost horizontal showers. However, we will now show that this approximation is questionable even for them as soon as one considers non-{\it standard} scenarios. We have investigated the effect of the regeneration for those detectors using the set of parametrizations  defined in Secs.~\ref{subsec:xs} and~\ref{subsec:elosses}. These expressions do not constitute an exhaustive list of all the alternative and exotic models that exist in the literature, but rather a selection of a few examples that allow us to point out the relevance of this process. The combinations of cross sections and tau energy losses should be chosen on basis of coherent PDFs but this coherency is hard to assert even inside the frame of the standard model. Therefore, we chose to investigate the regeneration effects for the 3$\otimes$3 possible combinations, which is sufficient to provide a qualitative answer to the question raised in this paper. 

As a starting point, one should however notice that the regeneration 
requires the $\tau$ to be converted back into a $\nu_\tau$ before
losing too much energy, and then this $\nu_\tau$ to undergo another
CC weak interaction to produce a $\tau$ again. Hence, one expects
that a higher energy loss or a lower cross section will reduce the
effect of the regeneration, while, on the other hand, a higher
cross section or a lower energy loss will enhance it. This is indeed
what we find. In the following, we focus on the particular
combinations for which the regeneration is not negligible anymore, and
compare them to the standard case, essentially on the basis of monoenergetic incident beams of $\nu_\tau$'s. This is because the $E^{-2}$ power-law spectrum usually assumed for cosmic neutrinos, although adequate to look at the global flux of emerging $\tau$ leptons, is completely dominated by the lowest energies, while the behavior at the highest energies may be important for harder fluxes of incident neutrinos.

\begin{figure}[!t]
    \centering
    \includegraphics[width=9cm]{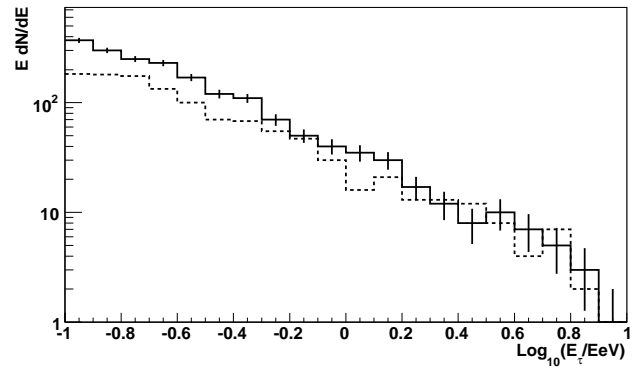}
    \caption{\small{Flux of emerging taus corresponding to incident tau neutrinos with an angle $\alpha$ up to 3$^{\circ}$ and $dN/dE = 4.6 \times 10^{7}$ E$_\nu^{-2}$~EeV$^{-1}$~sr$^{-1}$ using the combination of parameterizations $\sigma_{high}\otimes\beta_{low}$ as defined in Secs.~\ref{subsec:xs} and~\ref{subsec:elosses}. The regeneration is neglected in the dashed histogram and included in the solid one.}}
    \label{fig:HLE2_3degReg}
\end{figure}

In Fig.~\ref{fig:StdStdTau}-\ref{fig:LowHighTau}, we show the flux of $\tau$ leptons produced by incident neutrinos at given energies (0.3, 3 and 30 EeV) and with an angle $\alpha$ up to 3$^\circ$, for the following combinations of parametrizations: $\sigma_{std}\otimes\beta_{std}$ (the same as used
in Fig.~\ref{fig:StdE2_15Reg}), $\sigma_{high}\otimes\beta_{std}$, $\sigma_{std}\otimes\beta_{low}$ and  $\sigma_{high}\otimes\beta_{low}$. For a {\it standard} choice of parametrizations, one can see from Fig.~\ref{fig:StdStdTau} that the impact of regeneration is completely negligible at low neutrino energies. As the $\nu_\tau$ energy increases, however, the lowest-energy part of the $\tau$ spectrum starts to be significantly underestimated (15$\%$ less taus for 30 EeV incident neutrinos). This effect is washed out in the case of an incident neutrino flux $ dN/dE \propto E_\nu^{-2}$, for which the contribution of the highest-energy part of the neutrino spectrum is negligible; but it could affect the spectrum of emerging $\tau$'s if the neutrino flux were harder. For the other cases, the lowest-energy range of the $\tau$ spectrum becomes underestimated already for  incident $\nu_{\tau}$ with energy of 0.3 EeV. For the combination $\sigma_{high}\otimes\beta_{low}$, one loses about 70$\%$ of the $\tau$ leptons emerging from the Earth if the regeneration is neglected, even if assuming an incident flux of neutrinos $dN/dE \propto E_\nu^{-2}$ (Fig.~\ref{fig:HLE2_3degReg}).

The neutrino flux can be regenerated through the $\tau$ decay and the $\tau$ CC weak interaction. We present our results in a form that allows us to disentangle the two effects: the dotted histograms in
Fig.~\ref{fig:StdStdTau},\ref{fig:StdHighTau},\ref{fig:LowStdTau},\ref{fig:LowHighTau} account for regeneration through the weak interaction of the $\tau$ only, while the solid ones include both effects. As can be appreciated in the figures, the CC contribution is negligible at all energies in most scenarios. It only becomes important for $\nu_\tau$'s at the highest energies ($\sim 30$ EeV), in the energy range between 0.3 and 3.0 EeV of the emerging $\tau$ spectrum, if a {\it high} cross section is assumed (Figs. \ref{fig:StdHighTau} and \ref{fig:LowHighTau}). There, the effect is at the level of that from the regeneration through the $\tau$ decay.

\section{Conclusion}
\label{sec:conclusion}

In this paper, we have studied the mechanism of regeneration of the $\nu_\tau$ flux while crossing
the Earth, to investigate its effect on the flux of emerging $\tau$ leptons. Assuming a
detector with an energy threshold of 0.1 EeV and sensitive only to taus with an emerging
angle below 3$^\circ$, the effect is negligible for a flux of incident neutrinos $dN/dE \propto E_\nu^{-2}$ with the
standard values of cross sections and energy losses. But we have shown that this is not
valid for other assumptions on the detector performance, neither for less standard values
of the weak cross section or tau energy losses. On the one hand, neglecting the regeneration leads to about 30$\%$ underestimation on the integrated flux for a detector sensitive
up to 15$^\circ$. On the other hand, the underestimation of the $\tau$'s emerging from the Earth with an energy greater than 0.1 EeV and an emerging angle below 3$^\circ$ may reach 70$\%$ for the different values of $\sigma_{\nu_\tau}$ and $\beta_\tau$ studied in this paper. Moreover, the error made in the computation of the flux of emerging taus can be even larger for a harder flux of incident neutrinos since the contribution of the regeneration effect increases with the energy of the incident neutrino.

The simplification of neglecting the regeneration is thus only safe for particular values of
the physical properties playing a role on the propagation and specific detectors. It may
lead to a significant underestimation of the flux of emerging $\tau$'s when looking at nonstandard values of the weak cross section
or tau energy losses. Therefore, it should be carefully treated and accounted for when studying the systematics due to the uncertainties on those properties or while using the Earth-skimming
technique to test for instance higher weak cross-sections. Similarly, one should carefully
check the effect for the characteristics of the actual detector before neglecting the
regeneration.

\section*{Acknowledgments}
The authors would like to thank their colleagues of the Pierre Auger Collaboration for
useful discussions and suggestions. O. B. B. is supported by the Ministerio de
Educaci\'{o}n y Ciencia of Spain through the postdoctoral grant program. V.V.E.
acknowledges support from the European Community $\mathrm{6^{th}}$ Framework Program
through the Marie Curie Fellowship MEIF-CT-2005 025057.

\appendix

\section{Iterative solution of the transport equations}
 \label{Ap1}

We reproduce here a numerical method proposed in \cite{perrone,tseng},
generalizing it to take into account the NC weak interaction of the $\tau$ lepton.
We start from a solution of order 0 and iterate it up to get a
solution that satisfies the relation $F(\Phi(E,x),\partial_x
\Phi(E,x))=0$ as given by Eq.~\ref{eq:PhiNuTau} or~\ref{eq:PhiTau}.
For numerical reasons, it is better to work with numbers between 0
and 1 so that we write the equations in terms of the deviation to
the solution of order 0, $\Phi^0_{\nu_\tau}$, rather than in terms of
$\Phi_{\nu_\tau}$ itself. We therefore assume that the solution is
of the form
\begin{eqnarray*}
    \Phi_{\nu_\tau}(E,x) = \Phi_{\nu_\tau}^0(E)
    \exp{\bigg[-\frac{x}{\lambda_{\nu_\tau}(E)}(1-Z_{\nu_\tau}(E,x))\bigg]}
\end{eqnarray*}
where $\lambda_{\nu_\tau}(E)^{-1} =
\lambda_{\nu_\tau}^{CC}(E)^{-1}+\lambda_{\nu_\tau}^{NC}(E)^{-1}$
accounts for the possible sources of depletion of $\Phi_{\nu_\tau}$
at an energy $E$, and $Z_{\nu_\tau}(E,x)$ is a function to be
determined, which includes all contributions to the regeneration
of $\Phi_{\nu_\tau}$ at the same energy.

$Z_{\nu_\tau}(E,x)$ can be obtained iteratively by forcing it to
satisfy the transport equations, viewed as a recurrence relation
with the initial condition  $Z_{\nu_\tau}^0(E,x)=0$:
\begin{eqnarray}
    Z_{\nu_\tau}^{n+1}(E,x) = -x\frac{\partial Z_{\nu_\tau}^{n}(E,x)}{\partial x} +\hspace{3.5cm}  \nonumber\\
    +\rho\, \mathcal{N}_A\,\frac{\lambda_{\nu_\tau}(E)}{\Phi_{\nu_\tau}^n(E,x)}
    \int\frac{\mathrm{d}y}{1-y}\Phi_{\nu_\tau}^n\bigg(\frac{E}{1-y},x\bigg)
    \frac{\mathrm{d}\sigma_{\nu_\tau}^{NC}(y,\frac{E}{1-y})}{\mathrm{d}y} \nonumber\\
    +\rho\,\mathcal{N}_A\,\frac{\lambda_{\nu_\tau}(E)}{\Phi_{\nu_\tau}^n(E,x)}
    \int\frac{\mathrm{d}y}{1-y}\Phi_{\tau}^n\bigg(\frac{E}{1-y},x\bigg)
    \frac{\mathrm{d}\sigma_\tau^{CC}(y,\frac{E}{1-y})}{\mathrm{d}y} \hspace{0.1cm}\nonumber\\
    +\frac{\lambda_{\nu_\tau}(E)}{c\Phi_{\nu_\tau}^n(E,x)}
    \int\frac{\mathrm{d}y}{1-y}\Phi_{\tau}^n\bigg(\frac{E}{1-y},x\bigg)
    \frac{\mathrm{d}\Gamma(y,\frac{E}{1-y})}{\mathrm{d}y} \hspace{1.2cm}
\end{eqnarray}

As the two equations are coupled, it is necessary to know
$\Phi_\tau$ at a given order to proceed with the iteration at the
next order. Following the same philosophy as above, we assume that
the solution of Eq.\ref{eq:PhiTau} for $\Phi_\tau$ is of the form:
\begin{eqnarray*}
    \Phi_{\tau}(E,x) = \int_0^x \mathrm{d}u S(E(u-x),u) K(E,u,x)
\end{eqnarray*}
where the following notations have been introduced:
\begin{eqnarray*}
  S(E,x) &=& \rho\mathcal{N}_A \int \frac{\mathrm{d}y}{1-y}
  \Phi_{\nu_\tau}\bigg(\frac{E}{1-y},x\bigg)
  \frac{\mathrm{d}\sigma_{\nu_\tau}^{CC}(y,\frac{E}{1-y})}{\mathrm{d}y}\\
\end{eqnarray*}
\begin{eqnarray*}
  K(E,u,x)&=&\exp\bigg[\kappa(E,u,x)\cdot(1-Z_\tau(E,x)\\
  &+&Z_\tau(E(u-x),u))\bigg]\\
  \kappa(E,u,x)&=&\int_u^x \mathrm{d}v \bigg[\frac{\partial \gamma(E(v-x))}
  {\partial E}-\frac{1}{\lambda_\tau(E(v-x))}\bigg].
\end{eqnarray*}
and
$\lambda_\tau(E)^{-1}=\lambda_\tau^{dec}(E)^{-1}+\lambda_\tau^{CC}(E)^{-1}+\lambda_\tau^{NC}(E)^{-1}$.
The equation for  $Z_\tau^n$ then reads:
\begin{eqnarray*}
    &&Z_{\tau}^{n+1}(E,x) =
    \frac{1}{\Phi_\tau^n(E,x)}\\
    &&\times\int_0^x\mathrm{d}uS^{n+1}(E(u-x),u)K(E,u,x)\\
    &&\times\bigg[Z_\tau^n(E(u-x),u)-\kappa(E,u,x)\frac{\partial_x Z_{\tau}^{n}(E,x)}
    {(\partial_E\gamma(E)-\lambda_\tau^{-1}(E))}\bigg]\\
    &&-\frac{\rho\mathcal{N}_A\int\frac{\mathrm{d}y}{1-y}
    \Phi_{\tau}^n\bigg(\frac{E}{1-y},x\bigg)
    \frac{\mathrm{d}\sigma_{\tau}^{NC}(y,\frac{E}{1-y})}{\mathrm{d}y}}{(\partial_E\gamma(E)-\lambda_\tau^{-1}(E))
    \Phi_{\tau}^n(E,x)}
\end{eqnarray*}

again with $Z_\tau^0(E,x)=0$). Practically, the iteration can be stopped as soon as n=3.


\begin{thebibliography}{99}

\bibitem{nuoscil}  S.~Fukuda {\it et al.}  [Super-Kamiokande Collaboration],
  Phys.\ Rev.\ Lett.\  {\bf 86} (2001) 5656 [arXiv:hep-ex/0103033].

\bibitem{Learned:1994wg}
  J.~G.~Learned and S.~Pakvasa,
  Astropart.\ Phys.\  {\bf 3} (1995) 267
  [arXiv:hep-ph/9405296].

\bibitem{Andres:1999hm}
  E.~Andres {\it et al.},
  Astropart.\ Phys.\  {\bf 13} (2000) 1
  [arXiv:astro-ph/9906203].

\bibitem{ice3}  A.~Karle  [IceCube Collaboration],
  Nucl.\ Instrum.\ Meth.\  A {\bf 567} (2006) 438
  [arXiv:astro-ph/0608139].

\bibitem{antares}  J.~Carr  [ANTARES Collaboration],
  Nucl.\ Instrum.\ Meth.\  A {\bf 567} (2006) 428.

\bibitem{nemo}  I.~Amore [NEMO Collaboration],
  Int.\ J.\ Mod.\ Phys.\  A {\bf 22} (2007) 3509 [arXiv:0709.3991 [astro-ph].

\bibitem{nestor}  G.~Aggouras {\it et al.},
  Nucl.\ Instrum.\ Meth.\  A {\bf 552} (2005) 420.

\bibitem{Aynutdinov:2005sc}
  V.~Aynutdinov {\it et al.}  [Baikal Collaboration],
  Nucl.\ Instrum.\ Meth.\  A {\bf 567} (2006) 433
  [arXiv:astro-ph/0507709].

\bibitem{anita} S.~W.~Barwick {\it et al.}  [ANITA Collaboration],
  Phys.\ Rev.\ Lett.\  {\bf 96} (2006) 171101
  [arXiv:astro-ph/0512265].

\bibitem{rice} I.~Kravchenko {\it et al.},
  Phys.\ Rev.\  D {\bf 73} (2006) 082002
  [arXiv:astro-ph/0601148].

\bibitem{forte} N.~G.~Lehtinen {\it et al.}
Phys.\ Rev.\  D {\bf 69} (2004) 013008
  [arXiv:astro-ph/0309656].

\bibitem{glue} P.~W.~Gorham  {\it et al.},
  Phys.\ Rev.\ Lett.\  {\bf 93} (2004) 041101
  [arXiv:astro-ph/0310232].

\bibitem{hankins} T.~H.~Hankins  {\it et al.},
  MN-RAS {\bf 283} (1996) 1027

\bibitem{james}
  C.~W.~James and R.~J.~Protheroe,
  arXiv:0802.3562 [astro-ph].

\bibitem{swarup}
  G.~Swarup and S.~Panda,
  arXiv:0805.4304 [astro-ph].

\bibitem{pao} J.~Abraham {\it et al.}  [Pierre Auger Collaboration],
 Phys.\ Rev.\ Lett.\ {\bf 100} (2008) 211101
  [arXiv:0712.1909]

\bibitem{hires}   K.~Martens  [HiRes Collaboration],
  arXiv:0707.4417 [astro-ph].

\bibitem{cosmogenic} F.~W.~Stecker {\it et al.},
  Phys.\ Rev.\ Lett.\  {\bf 66} (1991) 2697
  [Erratum-ibid.\  {\bf 69} (1992) 2738].

\bibitem{beresmirnov}  V. S. Berezinsky and A. Yu. Smirnov, Astrophys. Space
  Science {\bf 32} (1975), 461;

\bibitem{capelle} K.~S.~Capelle {\it et al.},
  Astropart.\ Phys.\  {\bf 8} (1998) 321
  [arXiv:astro-ph/9801313].

\bibitem{antoine} A.~ Letessier-Selvon, AIP Conf.Proc. 566 (2000) 157-171, arXiv:0009444 [astro-ph].

\bibitem{fargion} D. Fargion, Astrophys.\ J.\ 570 (2002) 909 [arXiv:hep-ph/0002453] and references therein.

\bibitem{DRSS} S.~I.~Dutta, M.~H.~Reno, I.~Sarcevic and D.~Seckel, \\
  Phys.\ Rev.\ D {\bf 63} (2001) 094020
  [arXiv:hep-ph/0012350].

\bibitem{bertou} X. Bertou {\it et al.}, Astropart. Phys.  {\bf 17} (2002) 183 [arXiv:astro-ph/0104452].

\bibitem{feng} J.~L.~Feng {\it et al.},
  Phys.\ Rev.\ Lett.\  {\bf 88} (2002) 161102
  [arXiv:hep-ph/0105067].

\bibitem{bottai} S.~Bottai and S.~Giurgola,
  Astropart.\ Phys.\  {\bf 18} (2003) 539
  [arXiv:astro-ph/0205325].

\bibitem{tseng}J.~J.~Tseng, T.~W.~Yeh, H.~Athar, M.~A.~Huang, F.~F.~Lee and G.~L.~Lin,
  Phys.~\ Rev.~\ D {\bf 68} (2003) 063003
  [arXiv:astro-ph/0305507].

\bibitem{montaruli} E.~Bugaev, T.~Montaruli, Y.~Shlepin and I.~Sokalski,
  Astropart.\ Phys.\  {\bf 21} (2004) 491
  [arXiv:hep-ph/0312295].

\bibitem{aramo} C.~Aramo {\it et al.},
  Astropart.\ Phys.\  {\bf 23} (2005) 65
  [arXiv:astro-ph/0407638].

\bibitem{sarkar}  L.~A.~Anchordoqui, A.~M.~Cooper-Sarkar, D.~Hooper and S.~Sarkar, 
  Phys.\ Rev.\  D {\bf 74} (2006) 043008
  [arXiv:hep-ph/0605086].

\bibitem{parente} N.~Armesto, C.~Merino, G.~Parente and E.~Zas,
  Phys.\ Rev.\  D {\bf 77} (2008) 013001
  [arXiv:0709.4461 [hep-ph]].

\bibitem{irimia} S.~Palomares-Ruiz, A.~Irimia and T.~J.~Weiler,
  Phys.\ Rev.\  D {\bf 73} (2006) 083003
  [arXiv:astro-ph/0512231].

\bibitem{anchordoqui}
  L.~Anchordoqui, T.~Han, D.~Hooper and S.~Sarkar,
  Astropart.\ Phys.\  {\bf 25} (2006) 14
  [arXiv:hep-ph/0508312].

\bibitem{nous} O.~Blanch {\it et al.},  Phys.\ Rev.\  D {\bf 77} (2008) 103004 [arXiv:hep-ph/0802.1119].

\bibitem{perrone} V.~A.~Naumov and L.~Perrone,
  Astropart.\ Phys.\  {\bf 10} (1999) 239
  [arXiv:hep-ph/9804301].

\bibitem{reya} E.~Reya and J.~Rodiger,
  Phys.\ Rev.\  D {\bf 72} (2005) 053004
  [arXiv:hep-ph/0505218].

\bibitem{CGC}  L.~D.~McLerran and R.~Venugopalan,
  Phys.\ Rev.\  D {\bf 50} (1994) 2225
  [arXiv:hep-ph/9402335]; see also E. Iancu and R. Venugopalan, in
  {\it Quark and Gluon Plasma} (World Scientific, Singapore, 2004)
  Vol. 3, p. 249.

\bibitem{han}  see T.~Han and D.~Hooper,
  New J.\ Phys.\  {\bf 6} (2004) 150
  [arXiv:hep-ph/0408348] and references therein.

\bibitem{Dutta:2005yt}
  S.~I.~Dutta, Y.~Huang and M.~H.~Reno,
  Phys.\ Rev.\  D {\bf 72} (2005) 013005
  [arXiv:hep-ph/0504208].


\end{thebibliography}
\end{document}